\documentclass[a4paper]{article}
\usepackage{latexsym}
\begin{document}
\title{Some Remarks on Lower Bounds for\\ Queue Machines\\
(Preliminary Report)}
\author{Holger Petersen\thanks{Part of the research was done while the author was with the Universit{\" a}t Stuttgart.}\\
Reinsburgstr.~75\\
70197 Stuttgart
}

\maketitle

\begin{abstract}
We first give an improved lower bound for the deterministic
online-simulation of tapes or pushdown stores
by queues. Then we inspect some proofs in a classical work on queue
machines in the area of Formal Languages and outline why a main
argument in the proofs is incomplete. Based on descriptional 
complexity, we show the intuition behind the argument to be correct.
\end{abstract}

\newcommand{\qed}{$\Box$}
\newtheorem{lemma}{Lemma}
\newtheorem{theorem}{Theorem}
\newtheorem{corollary}{Corollary}
\newtheorem{observation}{Observation}
\newtheorem{claim}{Claim}

\section{Introduction}
A classical result states that a storage organized
as a queue can be the basis of universal computations.
Implicitly this has been shown by 
Post in \cite{Post43}, who proved the universality of normal
productions. A normal production $u\to v$ 
transforms a string $ux$ into $xv$. In order to obtain the equivalence
with general formal systems, nonterminal symbols are required.

This idea leading back to Post's work is used in \cite{Manna74},
where a Post machine is defined as an automaton
with a finite control and queue storage. 
Some articles are devoted to similar models in the area 
of Formal Languages. Vollmar \cite{Vollmar70} considers
automata with queue storage. He focusses on automata working
in real-time (the number of steps executed is equal to the length
of the input) accepting with empty storage. 
Separation of deterministic and nondeterministic variants of this
model are shown, the language classes are compared to the 
Chomsky Hierarchy, closure properties and decidability are considered.
We will discuss some lower bound proofs from this
investigation in Section~\ref{analysis}.
Another work on machines with queue storage in the area of Formal Languages is 
\cite{Brandenburg80}.

While the universality of queue storages was known for a long time, 
results concerning the complexity
of computations on such models were obtained much later.
In \cite{Li92}
mainly lower bounds for the mutual simulation of machines with a varying
number of queues, pushdowns, and tapes are shown.
The simulation of a pushdown by a queue requires $\Omega(n^{4/3}/\log n)$
steps in the deterministic and nondeterministic case
(here $n$ denotes the number of steps being simulated). 
An optimal quadratic lower bound holds for the deterministic 
simulation of a queue by a tape \cite{Li88ic}, while
a nondeterministic simulation is possible in time $O(n^{3/2}\log^{1/2} n)$
\cite{Li88jcss}. The lower bound for this simulation
is $\Omega(n^{4/3}/\log^{2/3} n)$ \cite{Li88ic}.

For linear tapes it is known that many storages can be simulated
efficiently on two storages. In the deterministic case 
the bound $O(n\log n)$ is due to Hennie and Stearns
\cite{Hennie66a}, while a linear-time solution exists for 
nondeterministic machines \cite{Book70}.
The latter result can be transferred to queue storages
\cite[Theorem~4.5]{Brandenburg80} and \cite[Theorem~4.2]{Li92}.
Li, Longpr{\' e} and Vitanyi left open the question 
whether a subquadratic solution is possible for queue storages
in the deterministic case. H{\"u}hne \cite{Huehne93} could show 
that machines with several linear storages can be simulated by
$k$ queues in time $O(n^{1+1/k})$. For online simulations
this solution is almost optimal since a lower bound 
$\Omega(n^{1+1/k}/\mbox{polylog } n)$ holds. 
Below we give an overview of lower and upper bounds 
for simulations by queue machines.

Deterministic simulations by one queue:
\begin{center}
\begin{tabular}{|l|c|c|}\hline
\parbox{22ex}{storage being simulated} & \parbox{24ex}{\hfil lower bound} & \parbox{22ex}{\hfil upper bound}\\\hline
one-turn   & $\Omega(n^{4/3}/\log n)$ & $O(n^{3/2})$\\
 pushdown                      & \cite[Theorem~3.2]{Li92} &  \cite[Theorem 3]{Petersen06}\\\hline
one or two pushdowns,      & $\Omega(n^{4/3}/\log n)$ & $O(n^2)$\\
one tape                   & \cite[Theorem~3.2]{Li92} & \\\hline
two queues& $\Omega(n^2)$  &  $O(n^2)$\\
         & \cite[Theorem~4.13]{Li92} &\\\hline
\end{tabular}
\end{center}

Deterministic simulations by $k$ queues:
\begin{center}
\begin{tabular}{|l|c|c|}\hline
\parbox{22ex}{storage being simulated} & \parbox{24ex}{\hfil lower bound (online) } & \parbox{22ex}{\hfil upper bound}\\\hline
$k+1$ queues & $\Omega(n^{1+1/k}/\log^{1/k} n)$ & $O(n^{1+1/k})$\\      
             & \cite[Theorem~4.2]{Huehne93} & \cite[Theorem~3.2]{Huehne93}\\\hline      
three pushdowns,  & $\Omega(n^{1+1/k}/\log^{1/k} n)$ & $O(n^{1+1/k})$\\      
two tapes              & Theorem~\ref{lowerqbypd} & \cite[Theorem~3.2]{Huehne93}\\\hline      
\end{tabular}
\end{center}

Nondeterministic simulations by one queue:
\begin{center}
\begin{tabular}{|l|c|c|}\hline
\parbox{22ex}{storage being simulated} & \parbox{24ex}{\hfil lower bound} & \parbox{22ex}{\hfil upper bound}\\\hline
two queues & $\Omega(n^2/\log^2n\log\log n)$ &  $O(n^2)$\\
          & \cite[Theorem~4.5]{Li92} &  \\\hline
\end{tabular}
\end{center}

Nondeterministic simulations by two queues:
\begin{center}
\begin{tabular}{|l|c|}\hline
\parbox{22ex}{storage being simulated} & \parbox{49ex}{\hfil upper bound}\\\hline
queues &  $O(n)$ \\
           &  \cite[Theorem~4.5]{Brandenburg80}, \cite[Theorem~4.2]{Li92}\\\hline
one pushdown &  $O(n \log n)$\\
           &  \cite{Rosenberg98}\\\hline
multi-dimensional & $O(n \log^2 n)$\\
tapes               &  \cite{Petersen06}\\
\hline
\end{tabular}
\end{center}

\section{Preliminaries}
Queue machines are usually defined in the same way as Turing machines,
the storage consisting of one or several queues instead of tapes 
\cite{Vollmar70,Li88ic,Li92,Huehne93}.
The machines have a separate one-way input tape and (determined by
a finite control) can do the following in one step:
\begin{itemize}
\item Read zero or one symbol from the input.
\item Pop zero or one symbol from the queue.
\item Push zero or one symbol onto the queue.
\item Change state.
\end{itemize}
Acceptance is indicated in different ways: By writing a 0 or a 1 
\cite{Li88ic} or by empty storage \cite{Vollmar70,Li92,Huehne93}.
A {\em simple buffer automaton} in the sense of \cite{Vollmar70}
is a queue machine accepting with empty storage and working in
real-time (one input symbol is read in each step). Notice
that in \cite{Vollmar70} the empty string is  not
accepted by definition.

The Post machine  from \cite{Manna74} is more restricted and has no 
separate input tape. The input is initially stored on the queue instead.
This model is surprisingly powerful and can accept languages
like $\{ a^nb^n \mid n\ge 0\}$ in linear time 
(the algorithm in \cite[Example~1-9]{Manna74} is quadratic).
Lower bounds on models with a separate input tape
carry over to this model, since a Post machine can be simulated
without any overhead. In a first stage the input tape
supplies symbols the Post machine would read from its queue.
When the input has been consumed, the simulator switches 
to the queue.

\section{A Lower Bound for Simulating Several Storages}
H{\" u}hne has shown in \cite[Corollary~4.4]{Huehne93}, 
that the online-simulation of $2(k+1)$ 
tapes or pushdown stores by $k$ queues requires
$\Omega(n^{1+1/k}/\log^{1/k} n)$ steps
and that the analogous 
simulation of two tapes or three pushdown stores requires  
$\Omega(n^{1+1/k}/\log^{1 + 2/k} n)$ time
\cite[Corollary~4.5]{Huehne93}.
The proof is based on 
lower bounds for the
computation of certain functions by machines with 
$k$ queues.
Our proof uses the same functions as the proof of 
H{\" u}hne, the difference being that we give specific
algorithms for computing the functions on the models
which are hard to simulate by machines with $k$ queues.
In contrast, H{\" u}hne's proof uses general simulation
results causing an additional overhead.

Function $F_k$ is defined via a machine with $k$ queues
running in linear time.
The relevant subset of possible input strings (on other
inputs the output is arbitrary) is defined as  
\begin{eqnarray*}
L_k & = & \{ x_{1,1}\cdots x_{1,f_1}\#\cdots\# x_{i,1}\cdots x_{i,f_i}
                             \#\cdots\# x_{k,1}\cdots x_{k,f_k}\$\\
    &   & \quad x_{1,f_1+1}\cdots x_{i,f_i+1}\cdots x_{k,f_k+1}\$\cdots\$\\
    &   & \quad x_{1,f_1+j}\cdots x_{i,f_i+j}\cdots x_{k,f_k+j}\$\cdots\$\\
    &   & \quad x_{1,f_1+m}\cdots x_{i,f_i+m}\cdots x_{k,f_k+m}\$\mid \\
    &   & \qquad\forall 1\le i\le k, 1\le j\le f_i+m: x_{i,j}\in\{ 0, 1\}\}.
\end{eqnarray*}

On an input from the set $L_k$ the $k$ queue machine
$M_k$ fills its $i$-th queue with $x_{i,1}\cdots x_{i,f_i}$ 
from the section of the input before the first \$ 
(in this phase the output is irrelevant).
After the first symbol \$ has been read, $M_k$ works in
$m$ rounds. In round $j$
machine $M_k$ reads the symbols of the $j$-th string between \$-symbols 
and appends sequentially the $x_{i,f_i+j}$ 
to queue $i$, while the first symbols are deleted and output. 
Every \$ is output directly. Consequently, the lengths of the
queues are kept constant after the first 
 \$ has been read, and the output is 
$x_{1,1}\cdots x_{k,1}\$\cdots\$ x_{1,j}\cdots x_{k,j}\$\cdots
\$x_{1,m}\cdots x_{k,m}$. Machine $M_k$ works in linear
time and computes $F_k$.
   
\begin{lemma}\label{functionq}
For every $k\ge 1$ the function $F_k$ defined above
can be coumputed by a deterministic Turing machine with one tape
and one pushdown store in linear time.
\end{lemma}
Proof. Turing machine $T_k$ computing $F_k$ uses  
$k$ tracks on the work-tape in order to store the $k$ strings
$M_k$ would keep on its queues. After $x_{i,1}\cdots x_{i,f_i}$
has been read and copied onto the $i$-th track,
$T_k$ returns to the first cell of the portion of the
tape storing this string. After reading the first \$,
machine $T_k$ starts to output the stored symbols
and replaces them with symbols just read. 
Additionally, they are marked as new. 
When the string on track $i$ has been output completely, 
$T_k$ returns to the first new symbol while copying the symbols read onto the
pushdown store, then moves its head to the position where
reading had been suspended,
and writes the contents of the pushdown store onto the following
cells of track $i$. The section read is marked as old and 
reading resumes at the original position.

The initial phase in which the string $x_{i,1}\cdots x_{i,f_i}$
is read can be completed in a number of steps proportional
to the length of the string before the first \$.
Copying the segment on track $i$ is possible in $O(f_i)$ steps,
after $f_i$ symbols have been output.\qed

We are now able to the improve the lower bound from
\cite{Huehne93}.

\begin{theorem}\label{lowerqbypd}
Every  online-simulation of one tape and one push\-down store,
 two tapes, or three pushdown stores by deterministic machines with
$k$ queues requires  
$\Omega(n^{1+1/k}/\log^{1/k} n)$ steps.
\end{theorem}
Proof. The two other models can simulate one tape and one pushdown store
in linear time. Therefore, it is sufficient to show the lower bound
for the latter model.

According to Lemma~\ref{functionq} the function $F_{k+1}$ can be computed
in linear time by one tape and one pushdown store.
The proof of Theorem~4.2 in \cite{Huehne93} shows, that every machine with
$k$ queues computing $F_{k+1}$ requires 
$\Omega(n^{1+1/k}/\log^{1/k} n)$ steps for input length 
$n$.\qed

\section{Analysis of Lower Bound Proofs for Queue Machines}\label{analysis}
In \cite{Vollmar70} several separation results related to the 
Chomsky Hierarchy are presented. 
At the heart of the proofs is an argument claiming a 
quadratic lower bound on the time required to accept a certain
language in a restricted way described below. 
We will review the proofs of these results.

A central claim of \cite{Vollmar70} with a detailed proof is the 
following:
\begin{theorem}[Satz~3.4~(b) of  \cite{Vollmar70}]
There are deterministic context-sen\-si\-tive, non-con\-textfree languages
that are not accepted by deterministic simple buffer automata.
\end{theorem}

Approximately 20 years later, Li, Longpr\'e, and Vit\'anyi obtained a
stronger result by showing that the following deterministic context-sensitive, 
non-context\-free language in Section~4.2 of \cite{Li92} requires almost quadratic 
time on non-deterministic queue machines:
\begin{eqnarray*}
Q & = & \{ a\&b_0b_1\cdots b_k\# b_0b_0b_1b_2b_1b_3\cdots b_{2i}b_{i}b_{2i+1}
\cdots b_{k-1}b_{(k-1)/2}b_{k}\\
& & \quad b_0b_{(k+1)/2}b_1b_2b_{(k+3)/2}b_3\\
& & \quad \cdots b_{2i\bmod{(k+1)}}b_ib_{(2i+1)\bmod{(k+1)}}
\cdots b_{k-1}b_{k}b_{k}\& a \mid\\
& &\quad b_i \in \$\{0, 1\}^*\$ \mbox{ for } 0 \le i \le k, \mbox{ $k$ is odd}, a\in \{0, 1\}^*\}
\end{eqnarray*}
(the language is renamed in order to disambiguate it from $L$ below).

The earlier claim from \cite{Vollmar70}
is nevertheless interesting, since it is based on the simple witness-language
$$L = \{ wvcvw \mid v \in \{ 0, 1 \}^+, w \in \{ a, b \}^+\}$$
with a much shorter proof.
Notice also that $L$ can be accepted in linear time by a one-queue machine
by storing $wv$ on the queue, moving the copy of 
$w$ to the end of the queue, and comparing the queue contents to the input. 
This language is thus much closer to the class of languages accepted by 
simple buffer automata than the complex language $Q$. For these reasons
a convincing proof based on $L$ would be interesting.

We outline the main steps in the proof from \cite{Vollmar70} assuming that the input
has the form $wvcvw$ and thus should be accepted:
\begin{enumerate}
\item A deterministic queue machine $B$ accepts $L$ in real-time.
\item\label{unique} For every first portion $wvc$ of the input, $B$ has to enter a 
unique configuration.
\item\label{linear} Using a counting argument, the length of the queue has to be 
proportional to $|wv|$ when $wvc$ has been read.
\item By choosing $w$ sufficiently long, not all of the queue contents
can be removed while reading  $vcv$ and $v$ can be arbitrarily long.
\item\label{problem} After having read the queue contents caused by $w$, 
machine $B$ necessarily compares the queue contents encoding the two copies of $v$.
\item For the comparison, the two encodings of $v$ have to be moved to the front of the queue. 
This implies a quadratic running time, contradicting the real-time restriction of $B$.
\end{enumerate}
In comparison to Satz~3.4~(b) of \cite{Vollmar70}, the proof implies a stronger lower bound
quadratic in $|v|$ under the assumption that the prefix $wvcv$ is read in real-time. 

Claim \ref{unique} follows by a ``cut-and-paste'' reasoning. Item \ref{linear} does not
hold for compressible strings. From item~\ref{problem} the proof is incomplete, even when 
$w$ is incompressible.

We are not able to give a more efficient general algorithm for 
$L$ reading the portion $wvcv$ of the input in real-time 
than suggested by the proof of \cite{Vollmar70}. In light of 
Lemma~\ref{lowerL} we conjecture that a quadratic lower bound 
does indeed hold.

Instead of improving the algorithm for $L$, 
we will first define a language $L'$ that satisfies the first items
listed above (including item~\ref{linear} for certain inputs), 
but can be accepted by a deterministic queue machine in a way
contradicting the claims starting with 
item~\ref{problem} above. In addition, we will characterize a subset $L''$ of $L$
based on compressible strings that can be accepted in real-time.

Let 
$$L' = \{ wvcv\pi(w) \mid v \in \{ 0, 1 \}^*, w \in \{ a, b \}^*, |w| = 2^{|v|}\}$$ 
where $\pi$ is a permutation of the symbols of a word $w = x_1x_2\cdots x_{2^k-1}x_{2^k}$
of length $2^k$ for $k \ge 0$ with $x_i \in \{ a, b \} \mbox{ for } 1\le i \le 2^k$
defined by
\begin{eqnarray*}
\pi(w) & = & x_1x_3\cdots x_{2^k-1}x_2x_6\cdots x_{2^k-2}
\cdots x_{2^m}\cdots x_{2^k-2^m}\cdots x_{2^k}
\end{eqnarray*}
Permutation $\pi$ can be defined arbitrarily on strings not having a length that is
a power of two.

\begin{observation}
Language $L'$ can be accepted by a deterministic queue machine $M$ that works in real-time on the prefix 
$wvcv$ and in linear overall time on accepted inputs.
\end{observation}
Proof. We make use of the fact that computations making a bounded number of steps between reading
consecutive input symbols can be converted into real-time computations (by compressing the storage).

The queue machine $M$ accepting $L'$ first stores a prefix of the input from 
$\{ a, b \}^*\{ 0, 1 \}^*c\{ 0, 1 \}^*$  onto the queue.
When reading the next symbol from $\{ a, b \}$, the machine starts to read one symbol $x$ from
the queue, compares it to the current input symbol, and rejects if the symbols are different. 
The next symbol on the queue is moved to the end of the queue.
Continuing this process, $M$ removes one out of two symbols from the queue until a symbol from 
$\{ 0, 1 \}$ is at the front of the queue. Then $M$ remembers this symbol 
in its finite control and drops it while keeping all symbols from $\{ 0, 1 \}$ until $c$ is read from the 
queue. The next symbol from $\{ 0, 1 \}$ after $c$ is compared with the symbol stored in the 
finite control rejecting the input if  the symbols are different. The following string from $\{ 0, 1 \}^*$
is cycled to the rear of the queue. If all stored symbols of $w$ and $v$ can
be removed in this way, $M$ accepts its input.

The symbols of a prefix $w\in \{ a, b \}^*$ of the input will be permuted according to $\pi$ on the queue
under the assumption that the length constraints are satisfied.
If the input has the form $wvcv'w'$ with $v,v' \in \{ 0, 1 \}^*$, $|v| = |v'|$, $w, w' \in \{ a, b \}^*$, and  
$2^{|v|}=|w| = |w'|$ then $M$ accepts if $v = v'$ and $\pi(w) = w'$. If the length constraints are not satisfied, 
$M$ rejects.

For the time bound notice that the steps of $M$ after having read $wvcv$ are determined by the size of the queue.
If we divide $M$'s computation into cycles that start with a queue storing a string from 
$\{ a, b \}^*\{ 0, 1 \}^*c\{ 0, 1 \}^*$, then the length of the queue at the start of cycle $i$ is $2^{k-i+1}+2(k-i+1)+1$, where 
$k= |v|$. 

On an accepted input (for which $k^2 = \log^2|w| = O(n)$) 
the total running time after having read $wvcv$ can be bounded by 
$$2+\sum_{i=1}^{k+1}2^{k-i+1}+2(k-i+1) + 1 = 2 + 2^{k+1}-1 + k^2 + 2k + 1 = O(n).$$

Since the first part  of the computation is real-time, we obtain the claimed time bound.
\qed

The second language exhibiting gaps in the proof of Satz~3.4~(b) from \cite{Vollmar70} is:
\begin{eqnarray*} L'' & = & \{ a^mvcva^m \mid m = n\cdot \lceil\log_2 n\rceil - 2^{\lceil\log_2 n\rceil} + 1 
\mbox{ for some $n\ge 1$},\\ 
  & & \qquad v \in \{0, 1\}^+, 2|v|+1 \le \lceil\log_2 n\rceil \}.
\end{eqnarray*}
Notice that the lengths of the blocks of $a$s are determined by a specific counting technique
and correspond to sequence A001855 from \cite{OEIS}.

Language $L''$ is a proper subset of $L$ and we will describe an algorithm processing 
{\em every} input in real-time.
\begin{observation}
Language $L''$ can be accepted by a deterministic queue machine $M$ that works in real-time.
\end{observation}
Proof. We will informally describe the operation of a deterministic queue machine $M$ with a
queue alphabet $\Sigma = \{ 0, 1, \$ \} \times \{ 0, 1, b, c, d \}$. We can assume that
$M$'s input is of the form $a^{m_1}v_1cv_2a^{m_2}$ with $m_1, m_2 \ge 0$ and $v_1, v_2 \in \{0, 1\}^+$
because $M$ can check this format with the help of its finite control.  

Since a prefix $a^m$ can be represented efficiently by its length in binary, the first phase of the 
recognition algorithm maintains a binary counter on the queue with the least significant bit at the front of
the queue. The digits of the counter are recorded in the first components of queue symbols, while
the second components are $b$ in this phase and will be ignored in the discussion. 
Machine $M$ works in cycles starting with a single separator symbol \$ on the queue. 
Initially $M$ is in a state that corresponds to a carry to the next significant digit.
If the first symbol on the queue is a separator symbol \$, then $M$ removes it and writes 
$0$ followed by \$ onto the queue. If symbol $1$ is at the front of the queue, it is replaced with a
$0$ at the rear of the queue and $M$ remains in the state corresponding to a carry. In the case 
of a $0$ at the front of the queue, $M$ replaces it with a $1$ and switches to a state that
moves symbols from the front to the rear of the queue until a \$ appears at the front. In parallel
to the activities described, $M$ reads an input symbol in every step. The first phase
ends when a symbol from $\{0, 1, c\}$ is read from the input tape. At this moment $M$ checks that the
symbol at the front of the queue is \$ and rejects if it is not not.

In the second phase, $M$ processes the middle portion of the input formed over the alphabet $\{0, 1, c\}$.
Each input symbol is encoded in the second component of a queue symbol while decrementing the counter
value encoded in the first components. Symbols are moved to the rear of the queue in this process. 
The second phase ends when the first $a$ after the $c$ is encountered.

In the third phase the string of $a$s after the middle portion is processed 
while the counter is continuously cycled and decremented. In order to reverse the increment process
in the first phase, the length of the counter is reduced by one bit if all digits are $0$ (this can be
checked with the help of the finite control during each cycle of the queue).
In a parallel process, $M$ compares bit by bit the second components of
symbols in the queue section storing the middle portion
of the input. One bit is recorded in the finite control while it is replaced by $d$ on the queue. The
recorded symbol is compared with the first symbol from $\{ 0, 1\}$ after the $c$. After the comparison this
symbol is also replaced by $d$. If a mismatch is found, $M$ rejects the input.
In order to compensate for reading the middle portion, $M$ determines whether the current number of digits 
of the counter equals $|v_1cv_2|$ and does a single extra cycle without decrementing the counter. 

Notice that with $k$ digits on the queue, $M$ performs at most $(k+1)2^k$ steps (\$ can appear 
in $(k+1)$ places) and this is the number of digits in the binary representation 
of all numbers with exactly $k+1$ digits. This explains the connection to 
sequence A001855 from \cite{OEIS} ($a(n)$ is the number of digits in the binary representation of 
all the numbers $1$ to $n-1$)
and we have chosen a closed form for the definition of $L''$.
\qed

We will now show a weaker lower bound than the one implied by the proof of Satz~3.4~(b) 
of \cite{Vollmar70}, but strong enough to establish the claimed separation from 
the deterministic context-sen\-si\-tive languages. The proof is based on 
descriptional complexity of strings, see \cite{Li97}. 
\begin{lemma}\label{lowerL}
Every deterministic simple buffer automata accepting $L$ and reading the portion $wvcv$ 
of accepted strings in real-time makes $\Omega(n^2/\log^3 n)$ steps in the worst case.
\end{lemma}
Proof. Let $B$ be an automaton with $s$ internal states and $q$ queue symbols as described 
in the lemma.

Let $u \in \{ 0, 1\}^*$ be an incompressible string 
with $|u| = n$ sufficiently large. We split
$u$ into words $v, w$ with $vw = u$ and $|v| = n / \log n$. 
Consider the uniquely determined accepting computation of $B$ on input $h(w)vcvh(w)$, where
$h(0) = a$ and $h(1) = b$. We first argue that for large $n$ the first 
copy of $v$ cannot influence $B$'s computation while reading the second
copy of $v$. This is clearly the case if the length of the queue
exceeds $|vcv| = 2|v|+1$ at the moment when $B$ starts to read the first copy of $v$.

The string $u$ can be reconstructed from the following information:
\begin{itemize}
\item A formal version of the algorithm described below ($O(1)$ bits).
\item An encoding of $B$ ($O(1)$ bits).
\item The value $n= |u|$ as a binary encoding in self-delimiting format ($2\log n$ bits).
\item The queue contents $z$ encoded in binary when $B$ starts to read the first copy 
 of $v$ ($|z|\cdot\log q$ bits).
\item $B$'s internal state when $B$ starts to read the first copy of $v$ ($O(1)$ bits).
\item The string $v$.
\end{itemize}
The decoder for $u$ runs $B$ starting from the internal state recorded on  
$ch(x)$ (temporarily assuming that $v$ is empty) for every binary $x$ of length $|w|$ with 
the initial queue contents $z$. If the decoder finds a suffix $ch(x)$ that $B$ 
accepts, it has determined $w = x$, since otherwise $B$ would accept an input 
$h(w)ch(w') \not\in L$ for $w \neq w'$. Now $u$ is determined as $vW$.

By incompressibility we have $c_1 + 2\log n + |z|\cdot\log q + n / \log n \ge n$ for a
constant $c_1$  and $|z| \ge (n - c_1 - n / \log n- 2\log n) /\log q$. It thus suffices
to choose a sufficiently large
$n$ such that $(n - c_1 - n / \log n- 2\log n) /\log q > 2n / \log n+1$

In the sequel we will distiguish the two copies of $v$ as $v_1$ and $v_2$.
We call the occurrence of $v_k$ in the input
$v_k$'s image of generation $0$. The queue section written while
$B$ reads $v_k$'s image of generation $i$ will be called  $v_k$'s image of generation $i+1$
for $i\ge 0$. Notice that at every point in time there are at most portions of two 
generations on the queue. The computation of $B$ is stopped 
in the process of reading images of generation $r$ 
as soon as the suffix of images of generation $r$ 
and the prefixes of images of generation $r+1$ together contain at most 
$|v| / (2\log q)$ symbols. This will happen eventually, since $B$ accepts with
empty queue. 

String $u$ can be reconstructed from the following information:
\begin{itemize}
\item A formal version of the algorithm described below ($O(1)$ bits).
\item An encoding of $B$ ($O(1)$ bits).
\item The value $n = |u|$ as a binary encoding in self-delimiting format ($2\log n$ bits).
\item String $w$ ($(1 - 1/\log n) n $ bits).
\item $B$'s internal state when $B$ is stopped ($O(1)$ bits).  
\item The lengths of the prefix and suffix of $v$'s images when 
$B$ is stopped ($O(\log n)$ bits).
\item The concatenation of prefix and suffix of $v$'s images when 
$B$ is stopped ($n / (2\log n)$ bits).
\item A list of $r+1$ records containing $B$'s states when it enters 
and leaves $v$'s images (empty for the last record) 
up to generation $r$ ($O(r)$ bits).
\item A list of the lengths of input segments read while 
 $v$'s images up to generation $r-1$ are written ($O(r\log n)$ bits). 
\end{itemize}
The decoder for $u$ systematically enumerates binary strings $y$ of length 
$n / \log n$. For every $y$ it sets up a string representing
the input $h(w)ycyh(w)$. Then it simulates $B$ until it enters the 
second copy of $y$ and checks that $B$'s state is consistent with the
state recorded. The decoder continues marking $y$'s images on the queue 
and checks state and position on the input when $B$ leaves the 
second copy of $y$. The decoder continues the simulation checking consistency 
every time $B$ enters or leaves one of $v$'s images until the list has been
exhausted. Then $B$ compares prefix and suffix of $y$'s images on 
the queue with the strings recorded. Since $B$ accepted from this 
configuration, the simulation can be terminated. 
The decoder proceeds to the
next $y$ whenever an inconsistency in the simulation is detected.

If the simulation succeeds, string $v$ (and thus $u$) has been determined, 
because a $y \neq v$ leading to a consistent simulation would imply
that $B$ accepts $h(w)vcyh(w) \not\in L$ and $h(w)ycvh(w) \not\in L$.

By incompressibility we have 
$$c_2 + 2\log n + (1 - 1/\log n) n + c_3\log n + n / (2\log n) + c_4r\cdot \log n \ge n$$
for constants $c_2, c_3, c_4$ and 
$$ r \ge (1/(c_4\log n)) \cdot (n / (2\log n) - c_2 - (c_3+2)\log n) = \Omega(n /\log^2 n).$$
Since $v$'s images have length $\Omega(n /\log n)$ until $B$ is stopped, we obtain 
a lower bound $\Omega(n^2 /\log^3 n)$ for the time bound of $B$.
\qed

Another result the proof of which refers to the incomplete argument 
in the proof of Satz~3.4~(b) of \cite{Vollmar70} is:
\begin{theorem}[Satz~3.5~(b) of \cite{Vollmar70}]
There exist deterministic context-free languages
 not accepted by deterministic simple buffer automata.
\end{theorem}
The witness language is
$ \{ wvcv^Rw^R \mid v \in \{ 0, 1 \}^+, w \in \{ a, b \}^+\}$.
The proof can be adapted from the one of Lemma~\ref{lowerL}, since the descriptional
complexity of a string and its mirror-image are equal up to a constant. 
The result also follows from the lower time bound $\Omega(n^{4/3}/\log n)$ 
on palindrome recognition by non-deterministic one-queue machines from \cite{Li92}.

Finally, the proof of the following non-closure property is affected by the incomplete
proof of Satz~3.4~(b):
\begin{theorem}[Satz~4.1~(a) of \cite{Vollmar70}]
The class of languages accepted by deterministic simple buffer automata is not closed under intersection.
\end{theorem}
This claim does not seem to follow from other known results, and thus 
the representation
\begin{eqnarray*}
L & = & \{ wv_1cv_2w \mid v_1, v_2 \in \{ 0, 1 \}^+, w \in \{ a, b \}^+\} \\
  &   & \quad \cap \quad \{ w_1vcvw_2 \mid v \in \{ 0, 1 \}^+, w_1, w_2 \in \{ a, b \}^+\} 
\end{eqnarray*}
together with Lemma~\ref{lowerL} now provides a proof.

\section{Discussion}
For the online simulation of two tapes or three pushdown stores by $k$ queues we could 
improve the lower bound
$\Omega(n^{1+1/k}/\log^{1+2/k} n)$ 
 to $\Omega(n^{1+1/k}/\log^{1/k} n)$. 
Then we investigated proofs in the classical paper \cite{Vollmar70} 
on queue machines. It turned out 
that a main argument is incomplete. We proved
a lower bound slightly weaker (by a $\mbox{polylog } n$ factor)
on the witness language $L$, but still sufficient to 
establish the results reported in \cite{Vollmar70}. It remains open 
whether the quadratic lower bound implied by the proof in \cite{Vollmar70} 
holds.


\end{document}